\documentclass[runningheads]{llncs}
\usepackage[T1]{fontenc}

\usepackage{graphicx}
\usepackage{amsmath}
\usepackage{graphicx}
\usepackage[colorlinks=true, allcolors=blue]{hyperref}
\usepackage{amssymb}
\usepackage{multirow}
\usepackage{booktabs}
\newcommand{\bs}{\boldsymbol}
\usepackage{bm}

\begin{document}
\title{Segmentation-guided MRI reconstruction for meaningfully diverse reconstructions}
\author{Jan Nikolas Morshuis \inst{1} \and Matthias Hein \inst{1} \and
Christian F. Baumgartner \inst{1,2}}
%
\authorrunning{JN Morshuis et al.}
\titlerunning{Segmentation-guided MRI reconstruction}

\institute{University of Tübingen \and University of Lucerne \\
\email{nikolas.morshuis@uni-tuebingen.de}}
\maketitle
\begin{abstract} 

Inverse problems, such as accelerated MRI reconstruction, are ill-posed and an infinite amount of possible and plausible solutions exist. This may not only lead to uncertainty in the reconstructed image but also in downstream tasks such as semantic segmentation. This uncertainty, however, is mostly not analyzed in the literature, even though probabilistic reconstruction models are commonly used. These models can be prone to ignore plausible but unlikely solutions like rare pathologies. Building on MRI reconstruction approaches based on diffusion models, we add guidance to the diffusion process during inference, generating two meaningfully diverse reconstructions corresponding to an upper and lower bound segmentation. The reconstruction uncertainty can then be quantified by the difference between these bounds, which we coin the 'uncertainty boundary'. We analyzed the behavior of the upper and lower bound segmentations for a wide range of acceleration factors and found the uncertainty boundary to be both more reliable and more accurate compared to repeated sampling. Code is available at \url{https://github.com/NikolasMorshuis/SGR}.

\end{abstract}
\section{Introduction} 

Accelerated MRI reconstruction has substantially improved in recent years through the use of machine learning methods. Currently, diffusion models (DMs) \cite{ho2020denoising,song2020score} achieve state-of-the-art MRI reconstruction scores \cite{chung2024decomposed,jalal2021robust,chung_score_based}. An often neglected area in the literature is the inherent uncertainty of the reconstruction process. Accelerated MRI reconstruction is an ill-posed problem and infinite plausible reconstructions exist for an undersampled MRI image. However, in the majority of recent work only a single point estimate (a single reconstruction) of the solution space is provided. This common practice is problematic in safety-critical applications like medical imaging, as it does not accurately reflect the inherent uncertainty and no estimation of the confidence of the solution can be provided.

\begin{figure}
    \centering
    \includegraphics[width=0.95\textwidth]{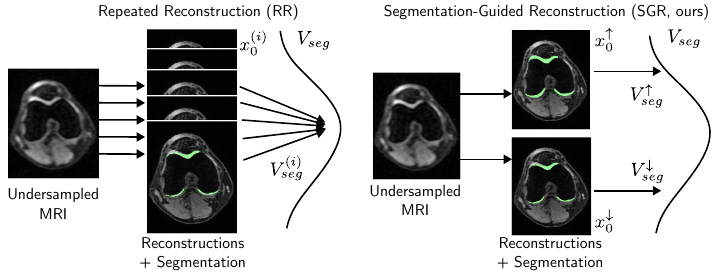}
    \caption{\textbf{Motivation.} Repeated reconstruction tends to create MRI reconstructions that are conceptually similar and not meaningfully diverse. Similar results are also observed in downstream tasks like segmentation in our experiments. Our method in contrast only reconstructs two images, an upper bound ($x_0^{\uparrow}$) and a lower bound ($x_0^{\downarrow}$) reconstruction, corresponding to an upper ($V_{seg}^{\uparrow}$) and lower bound ($V_{seg}^{\downarrow}$) on the segmentation volume, giving an intuitive understanding of the segmentation uncertainty.}
    \label{fig:concept_figure}
\end{figure}

A common approach to quantify the reconstruction uncertainty is to reconstruct a large number of samples and analyze their pixel-wise variance. This kind of uncertainty prediction has been investigated with probabilistic networks \cite{paul_uncertainty,cheung2024metricguided}, Monte Carlo dropout \cite{dropout_rec} ensembling \cite{küstner_unc} or DMs \cite{chung_score_based,jalal2021robust,mri_reconstruction_miccai}.

Recently however, \cite{cohen2024posterior} has shown that Repeated Reconstruction (RR) in natural image reconstruction with DMs leads to similar images which display a lack of meaningful diversity. Even though the diffusion model is trained to model the posterior distribution accurately, the heavy tail of the true posterior distribution is often largely ignored in practice and plausible yet unlikely solutions with different semantic meaning are not found. This is a problem for medical imaging, where certain pathologies might be rare and unlikely. However, it is important to recover these cases precisely and include them as possible solutions when reconstructing multiple samples for uncertainty prediction.

The definition of meaningful diversity can depend on the downstream task and might be different for e.g. pathology detection (a pathology exists or not) or semantic segmentation (the segmented tissue might be smaller or larger). In this work we focus on diversity of semantic segmentation, as it is an important task in clinical practice and the results are more easily interpretable for non-experts.

In this paper, instead of random sampling, we propose to reconstruct two MRI-images per segmentation class corresponding to the upper and lower bound segmentation volume. To this end, we introduce a novel method called Segmentation Guided Reconstruction (SGR). SGR builds on DM-based MRI reconstruction. However, rather than standard sampling, we guide the reconstruction process with the gradient of segmentation losses that either penalize small segmentation volumes or large segmentation volumes. The resulting lower and upper bound reconstructions and segmentations are an intuitive way of understanding the segmentation uncertainty which arises due to the ill-posed nature of the reconstruction problem. See Fig. \ref{fig:concept_figure} for a visual explanation of the motivation. 
In summary, our contributions are:

\begin{itemize}
    \item We introduce a novel method Segmentation-Guided Reconstruction (SGR) to guide the reconstruction process of undersampled MRI data, leading to  meaningfully diverse reconstructions.
    \item We show that the diverse reconstructions lead to diverse segmentations that correspond to an upper and lower bound on the segmentation. We introduce the concept of 'uncertainty boundary' for the volume between both bounds.
    \item We show that our method captures the inherent uncertainty more faithfully than the commonly used Repeated Reconstruction (RR), potentially leading to safer MRI reconstruction methods
\end{itemize}

\section{Related Work}

Diffusion based inverse problem solvers have seen large success in MRI reconstruction \cite{jalal2021robust,chung_score_based,mri_reconstruction_miccai,song2022solving}, achieving state-of-the-art reconstruction scores \cite{jalal2021robust,chung_score_based,chung2024decomposed}. In contrast to supervised learning-based methods \cite{sriram2020end,zbontar2018fastmri}, trained DMs can be used for several acceleration factors. Recent reconstruction methods \cite{chung2024decomposed} allow for more than $80 \times$ faster inference time, reducing the computational burden for RR. With RR it is possible to reconstruct several distinct images, allowing to estimate the reconstruction uncertainty by analyzing the variance of the voxel values. However, the distinct images do not necessarily display meaningful diversity \cite{cohen2024posterior}.

While the reconstruction and segmentation problem are mostly tackled separately, some work exists aiming to improve segmentation results by combining reconstruction and segmentation in an end-to-end fashion \cite{pmlr-v121-caliva20a,RecSeg,segmentation_aware_recon,wu2023learning,bioengineering_k2s}. Wu et al. \cite{wu2023learning} show that fine-tuning a pre-trained reconstruction network with a task-specific network head leads to improved segmentation results. Similarly, the winning method of the K2S challenge \cite{bioengineering_k2s} also first trains a reconstruction network before fine-tuning the network for segmentation. These approaches, however, lead to degraded reconstructions. Moreover, all of these methods lack an analysis of the effects of the inherent reconstruction uncertainty on segmentation.

The literature on reconstruction uncertainty propagation to downstream tasks is relatively sparse and most methods utilize repeated sampling of probabilistic networks that might not offer meaningful diversity \cite{cohen2024posterior}. Feiner et al.~\cite{Feiner_2023} propose to utilize Monte Carlo sampling to estimate the reconstruction uncertainty that can then be propagated to classification tasks. Cheung et al.~\cite{cheung2024metricguided} use repeated sampling of CT reconstructions to obtain a distribution of several downstream metrics, which they use to calculate statistically valid prediction sets using conformal prediction. 
Fischer et al. \cite{paul_uncertainty} are the first to consider the propagation of the aleatoric uncertainty due to the reconstruction process in the context of segmentation. By repeated sampling of a probabilistic reconstruction network and subsequent segmentation, they were able to quantify the variance of the segmentation at pixel-level. However, data-consistency was not enforced and no upper or lower bound of the segmentation has been estimated.

\section{Method}

%
\subsection{Background diffusion models and MRI-reconstruction}

In the inverse problem of MRI reconstruction, we aim to reconstruct a plausible image $\bs{x}_0$ that is coherent with the measurement data $\bs{y}$ when passing through a forward operator (i.e. Fourier Transform + Masking) $\bs{A}$, such that the following equation holds:
\begin{equation}
    \bm{y} = \bs{A} \bs{x}_0^{(i)}
    \label{eq:inverse_problem} \ \text{.}
\end{equation}

The problem is typically ill-posed and infinite $\bs{x}_0^{(i)}$ exist that are possible solutions for Eq. \ref{eq:inverse_problem}. In recent years DMs have shown to perform exceptionally well on the MRI-reconstruction tasks, being able to sample a set of distinct solutions $\bs{x}_0^{(i)}$ from the posterior.

DMs generate samples from the data distribution $p_{data}(\bs{x}_0)$ by reversing a diffusion process from timestep $t=0$ to $t=T$ described by a conditional density

\begin{equation}
    q(\bs{x}_t|\bs{x}_0) = \mathcal{N}(\bs{x}_t|\sqrt{\bar{\alpha_t}} \bs{x}_0, (1-\bar{\alpha_t})\bs{I})
\end{equation}
with a pre-defined defined $\bar{\alpha_t}$ schedule.

A DM can then be trained by predicting the noise through epsilon matching:
\begin{equation}
    \min\limits_{\theta}\mathbb{E}_{\bs{x}_t \sim q(\bs{x}_t|\bs{x}_0), \bs{x}_0 \sim p_{data}(\bs{x}_0), \bs{\epsilon} \sim \mathcal{N}(0,\bs{I})} (||\bs{\epsilon}_{\theta}^{(t)}(\bs{x}_t) - \bs{\epsilon}||_2^2) \ \text{.}
\end{equation}
The learned $\bs{\epsilon}$-matching function can then be used to predict the clean image $\hat{\bs{x}}_{0|t}$ at every step $t$ of the diffusion process using Tweedie denoising \cite{efron2011tweedie}:

\begin{equation}
    \hat{\bs{x}}_{0|t} = (\bs{x}_t - \sqrt{1-\bar{\alpha_t}} \bs{\epsilon}_{\theta}^{(t)}(\bs{x}_t)) / \sqrt{\bar{\alpha_t}} \text{.}
\end{equation}

Recently, Chung et al. \cite{chung2024decomposed} introduced a DM based reconstruction method that utilizes conjugate gradient (CG) optimization: During the CG-optimization process, starting from $\hat{\bs{x}}_{0|t}$ we search a $\hat{\bs{x}}_{0|t}'$ that minimizes the distance $\bs{A}^*\bs{A} \hat{\bs{x}}_{0|t}' - \bs{A}^*\bs{y}$ with $\bs{A}^*$ being the backward operator mapping from the measurement space back to the image space:

\begin{equation}
\label{eq:conjugate_gradient}
    \hat{\bs{x}}_{0|t}' = CG(\bs{A}^{*}\bs{A}, \bs{A}^*\bs{y}, \hat{\bs{x}}_{0|t}) \text{.}
\end{equation}
The update rule using the DDIM method \cite{song2020denoising} is then given as follows:
\begin{equation}
    \bs{x}_{t-1} = \sqrt{\bar{\alpha}_{t-1}} \hat{\bs{x}}_{0|t}' + \sqrt{1-\bar{\alpha}_{t-1} - \eta^2 \tilde{\beta}_t^2} \bs{\epsilon}_{\theta}^{(t)}(x_t) + \eta \tilde{\beta}_t \bs{\epsilon} \text{.}
    \label{eq:ddim_update}
\end{equation}
We modify the algorithm from \cite{chung2024decomposed} slightly to achieve full data-consistency. For the last diffusion step $t=1 \rightarrow t=0$, we replace Eq. \ref{eq:conjugate_gradient} and make sure that the reconstructed image $\bs{x}_0$ is fully data-consistent using $\bs{y}$ and Eq. \ref{eq:inverse_problem}.

\subsection{Segmentation Guidance for diverse sampling}

\begin{figure}
    \centering
    \includegraphics[width=0.95\textwidth]{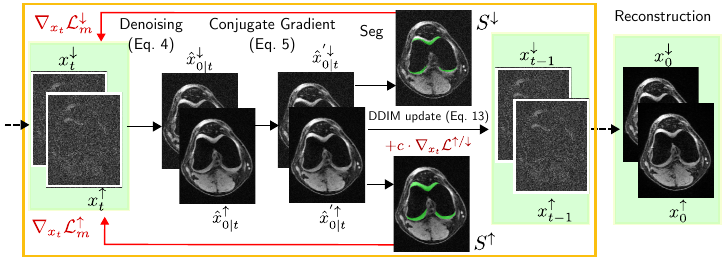}
    \caption{\textbf{Method explanation}. For each inverse diffusion step $t$, we calculate $\nabla_{x_t} \mathcal{L}^{\uparrow /\downarrow}$ and include the gradient in the calculation of $x_{t-1}$, in order to increase (decrease) the segmentation volume and to get an upper- and lower-bound segmentation ($S^{\uparrow}$ and $S^{\downarrow}$).}
    \label{fig:method_fig}
\end{figure}

In order to generate a set of meaningfully diverse solutions $\bs{x}_0^{(i)}$, we reconstruct two images per segmentation class ($\bs{x}_0^{\uparrow}$ and $\bs{x}_0^{\downarrow}$) with corresponding upper and a lower bound segmentations ($S^{\uparrow}$ and $S^{\downarrow}$) as well as segmentation volumes ($V_{seg}^{\uparrow}$ and $V_{seg}^{\downarrow}$). The volume between $S^{\downarrow}$ and $S^{\uparrow}$ is the uncertainty boundary $S_{unc}$. We guide the reconstruction process of the diffusion model using segmentation guidance. To achieve this, we propose two loss-functions $\mathcal{L}^{\uparrow}$ and $\mathcal{L}^{\downarrow}$ that are suitable for segmentation-based model guidance.

Let $p_{\phi}^c(\bs{x}_{0,jk})$ be the predicted probability of pixel $(j,k)$ in the image $\bs{x}_0$ belonging to the segmentation class $c \in \mathcal{C}$, and let $p_{\phi}$ be a segmentation network. We abbreviate the term as $\bs{p}_{jk}^{(c)}$. The volume of $c$ can be calculated by summing the number of pixels that belong to $c$ multiplied by voxel-volume $V_{voxel}$:

\begin{equation}
    V_c(\bs{x}_0) = \sum_{j,k} (\mathbb{I}_c(\bs{p}_{jk}^{(c)})) \cdot V_{voxel} \text{,}
\end{equation}
where $\mathbb{I}_c$ is an indicator function

\begin{equation}
    \mathbb{I}_c =     
    \begin{cases}
      1 & \text{if  } \bs{p}_{jk}^{(c)} > \bs{p}_{jk}^{(c')} \text{ } \forall c' \in \mathcal{C}, c' \neq c \\
      0 & $\text{otherwise}$
    \end{cases} 
    \text{.}
\end{equation}

During optimization, we consider only one class at a time, such that we can simplify the objective to a binary segmentation class. We build our guidance-loss based on the widely-used Binary-Cross-Entropy loss:

\begin{equation}
 \mathcal{L}_{BCE}(\bs{y}, \bs{p}, c) = - \sum_{j,k} \Bigl( \bs{y}_{jk}  \log(\bs{p}_{jk}^{(c)}) + (1-\bs{y}_{jk}) \log(1-\bs{p}_{jk}^{(c)}) \Bigr) \text{.}
    \label{eq:binary_ce}
\end{equation}

We can maximize the optimization objective $\mathcal{L}^{\uparrow}$ to find the upper bound segmentation, as well as $\mathcal{L}^{\downarrow}$ to find the lower bound segmentation:

\begin{equation}
    \mathcal{L}^{\uparrow} = \mathcal{L}_{BCE}(\bs{0}, \bs{p}, c) = - \sum_{j,k} \log(1-\bs{p}_{jk}^{(c)}) \text{,}
\end{equation}

\begin{equation}
    \mathcal{L}^{\downarrow} = \mathcal{L}_{BCE}(\bs{1}, \bs{p}, c) = - \sum_{j,k} \log(\bs{p}_{jk}^{(c)}) \text{.}
\end{equation}

The binary cross-entropy loss is unbounded, meaning that pixel probabilities are further optimized, even if the target class has already been reached. Following Croce et al. \cite{naman_robust_seg} we can avoid this behavior by using a masked cross-entropy loss, which excludes pixels already segmented as the target class from the optimization process. The optimization objectives can therefore be written as follows:

\begin{equation}
    \mathcal{L}_{m}^{\downarrow} = \mathbb{I}_c \mathcal{L}^{\downarrow}; \quad \mathcal{L}_{m}^{\uparrow} = (1-\mathbb{I}_c) \mathcal{L}^{\uparrow} \text{.}
\end{equation}

\begin{figure}[t!]
    \centering
    \includegraphics[width=0.95\textwidth]{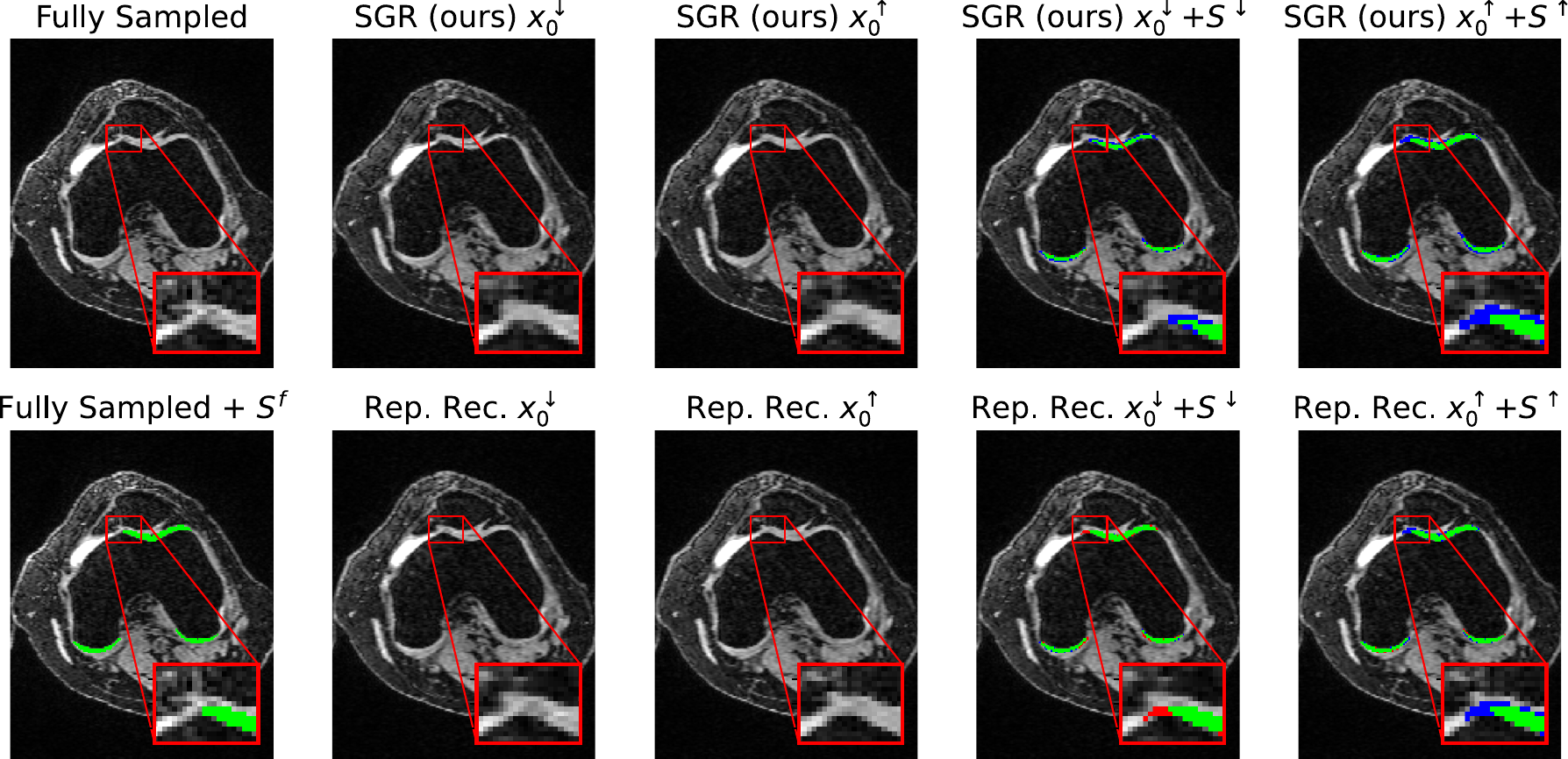}
    \caption{\textbf{Sampling vs. adversarial guidance.} Examples of lower- ($x_0^{\downarrow}$) and upper-bound ($x_0^{\uparrow}$) reconstructions (16x acc.) and segmentations ($S^{\downarrow}$, $S^{\uparrow}$) using our SGR method and the standard RR method. ($S^{f}$ is the segmentation of the fully-sampled image. Green: $S^{f} \& S^{\uparrow}$ or $S^{f} \& S^{\downarrow}$, Blue: $S^{f} > S^{\downarrow}$ or $S^{f} < S^{\uparrow}$, Red: $S^{f} < S^{\downarrow}$ or $S^{f} > S^{\uparrow}$)}
    \label{fig:method_example}
\end{figure}

The derived optimization objective can be used during the reverse diffusion process. We therefore calculate the gradient of $\mathcal{L}_m^{\uparrow }$ (or $\mathcal{L}_m^{\downarrow}$, respectively) with respect to $x_t$ at every step $t$, to find the perturbation necessary to increase the optimization loss. We add the gradient $\nabla_{x_t} \mathcal{L}_{m}^{\uparrow}$ (or $ \nabla_{x_t} \mathcal{L}_{m}^{\downarrow}$, respectively) during the DDIM update in Eq. \ref{eq:ddim_update}:

\begin{equation}
    \bs{x}_{t-1} = \sqrt{\bar{\alpha}_{t-1}} \hat{\bs{x}}_{0|t}' + \sqrt{1-\bar{\alpha}_{t-1} - \eta^2 \tilde{\beta}_t^2} (\bs{\epsilon}_{\theta}^{(t)}(x_t)+ \gamma \nabla_{\bs{x}_t}\mathcal{L}_{m}^{\uparrow / \downarrow}) + \eta \tilde{\beta}_t \bs{\epsilon} \text{,}
\end{equation}
where the parameter $\gamma$ makes sure that the $l_2$ norm of $\nabla_{x_t} \mathcal{L}_{m}^{\uparrow / \downarrow}$ stays small with respect to $\bs{\epsilon}_{\theta}$. We set $\gamma$ to the following value:

\begin{equation}
    \gamma =     
    \begin{cases}
      b \cdot ||\bs{\epsilon}_{\theta}^{(t)}||_2 / ||\nabla_{\bs{x}_t} \mathcal{L}_{m}^{\uparrow / \downarrow}||_2 & \text{if $||\nabla_{\bs{x}_t} \mathcal{L}_{m}^{\uparrow / \downarrow}||_2 > b ||\bs{\epsilon}_{\theta}^{(t)}||_2$} \\
      1 & \text{otherwise} \text{,}
    \end{cases} 
\end{equation}
where $b$ is a small constant that we set to $b=0.005$ in our experiments. An ablation study on the effects of the parameter value can be found in the Appendix. A summary of our method is shown in Fig. \ref{fig:method_fig}.

For the baseline RR, we sample 16 images per slice and select $x_0^{\uparrow}$ and $x_0^{\downarrow}$ for every segmentation class as the reconstruction leading to the highest segmentation volume ($V_{seg}^{\uparrow}$) and lowest segmentation volume ($V_{seg}^{\downarrow}$), respectively.

\begin{figure}[t!]
\centering
\includegraphics[width=0.95\textwidth]{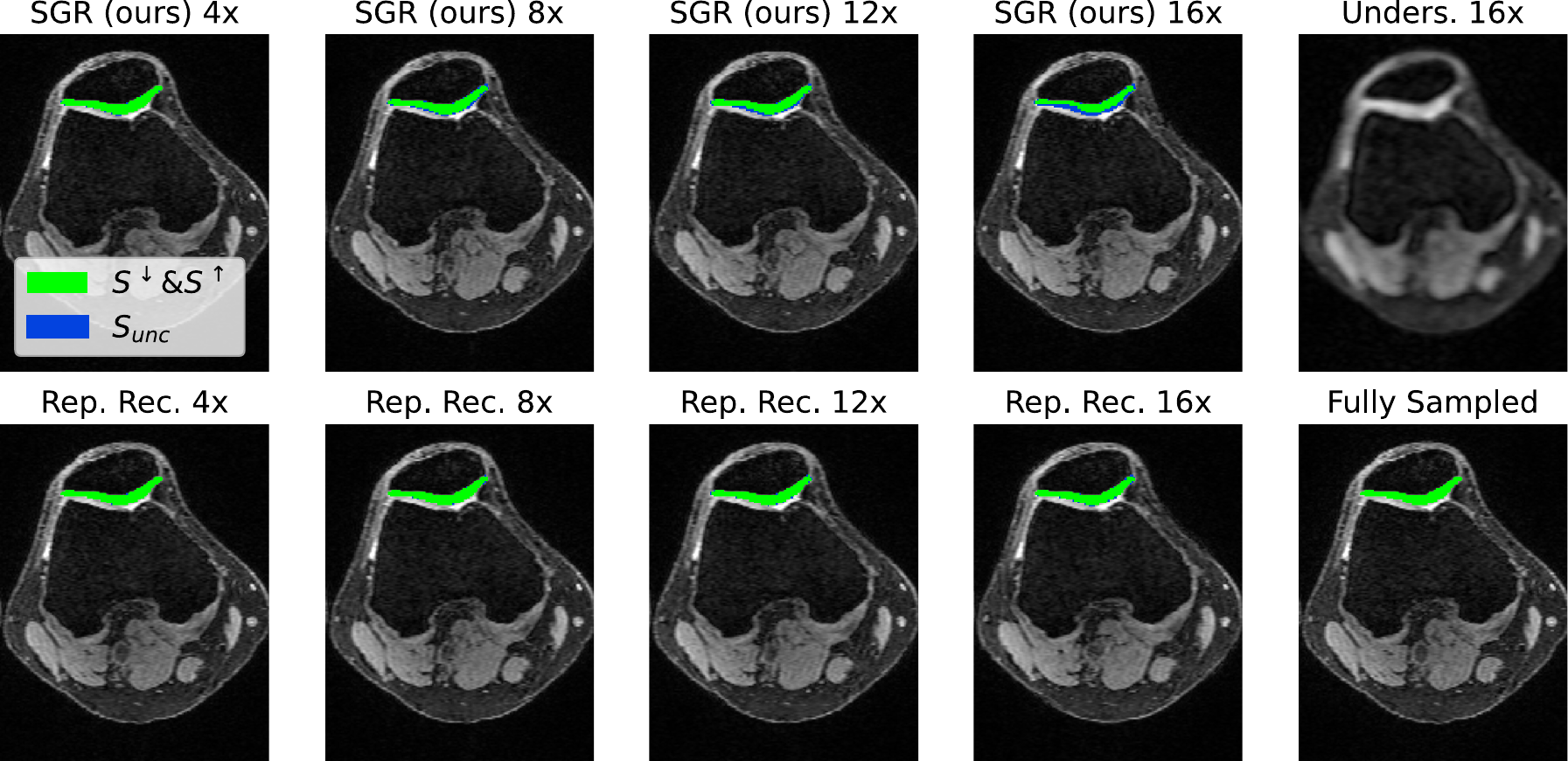}
\caption{\textbf{Analysis of different acceleration factors.} For RR, we see that the uncertainty does not increase, even though very high accelerations are tested. Our method, in contrast, generates more reliable uncertainty boundaries.}
\label{fig:different_accelerations}
\end{figure}

\subsection{Data and models}
We train and test our method on the publicly available SKM-TEA \cite{desai2021skm} dataset and focus only on the $E1$-echoes. As the spacing of the SKM-TEA data is very fine-grained (0.3125 mm in the axial ($=$sampling)  direction), neighboring images tend to be similar. We therefore save computational resources during testing by only considering every 8th slice in the relevant regions, corresponding to a spacing of 2.5mm between the slices. In total we analyze $869$ slices for each acceleration factor and each method. 

All experiments are based on the same diffusion model. We train the diffusion model on the fully-sampled images of the SKM-TEA data. We normalize the data by dividing by the 0.99 quantile of every slice. The diffusion model is based on the one introduced in  \cite{dhariwal2021diffusion}. We train the model for 1 million steps using a batch-size of 2 on a 3080TI GPU. For segmentation, we train a U-Net on the fully-sampled MRI-data in axial direction. Due to the strong class-imbalance on the axial imaging plane, where only a minority of slices show elements of tibial cartilage or meniscus as these parts tend to be relatively flat, we perform a foreground oversampling of 40\%. To reduce overfitting, we apply mirroring and Gaussian Noise augmentations during training.

\begin{figure}[t]
\centering
\includegraphics[width=0.99\textwidth]{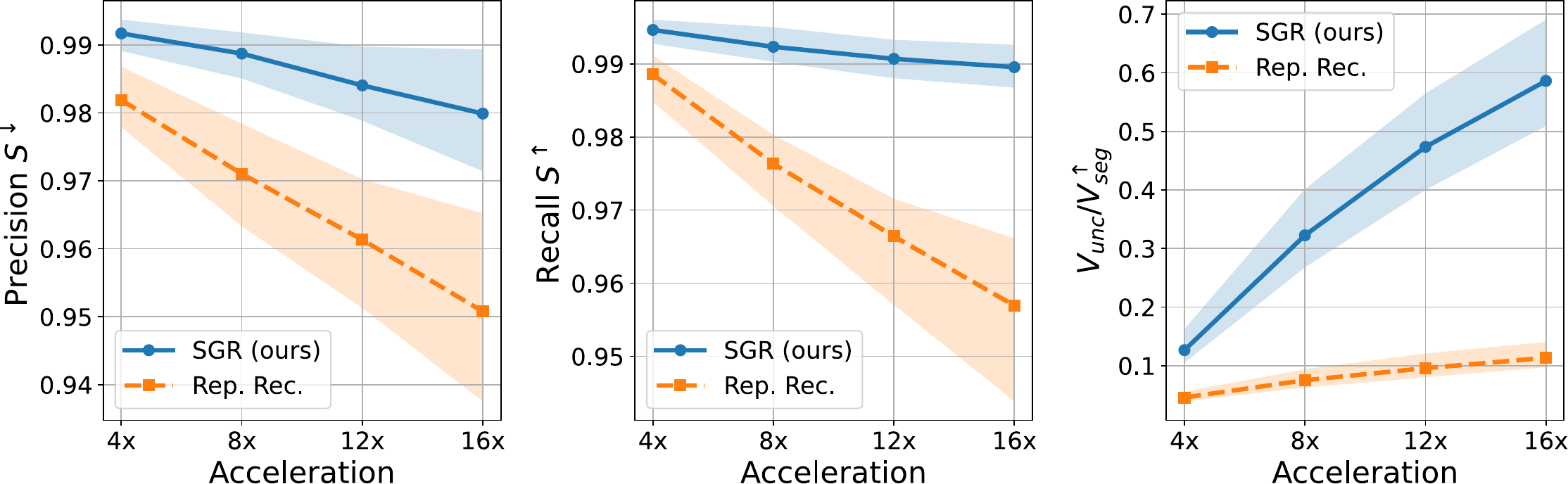}
\caption{\textbf{Results.} Median score and percentile boundaries (25\%, 75\%) for different acceleration factors. Note that the upper and lower bound segmentation of our method are mostly correctly capturing the inherent uncertainty even for high acceleration factors: we have high precision (small amount of false-positives in $S^{\downarrow}$) and recall (small amount of false-negatives in $S^{\uparrow}$). The ratio of uncertain volume $V_{unc}=V_{seg}^{\uparrow}-V_{seg}^{\downarrow}$ over $V_{seg}^{\uparrow}$ also increases, better reflecting the true underlying uncertainty.} 
\label{fig:ResultPlots}
\end{figure}

\section{Experiments and results}

We perform experiments on 4 acceleration factors: $4\times$, $8\times$, $12\times$, and $16\times$, analyzing the behavior of the uncertainty boundaries of SGR and RR. We assess the quality of the generated segmentations $S^{\downarrow}$ and $S^{\uparrow}$ by calculating the precision for $S^{\downarrow}$, recall for $S^{\uparrow}$ and the uncertainty volume $V_{unc}$. We focus on precision on $S^{\downarrow}$, as we want to avoid overestimating the segmentation and creating false-positives, and measure recall for $S^{\uparrow}$, because we want to avoid segmentations containing false-negatives. An example image where RR overestimates $S^{\downarrow}$ while our method predicts accurate boundaries is shown in Fig. \ref{fig:method_example}. A similar observation can be found in Fig. \ref{fig:ResultPlots}, where our method sustains high precision and recall values, while these scores degrade with higher acceleration for the baseline method RR. Reconstruction quality as measured by SSIM and PSNR remains very high for SGR compared to RR, as shown in Tab. \ref{tab:results}.

In Fig. \ref{fig:different_accelerations}, we see that the uncertainty boundary increases with higher acceleration for our method, while it stays approximately constant for the baseline method RR. This result is also shown in Fig. \ref{fig:ResultPlots}, where the ratio $V_{unc}/V_{seg}^{\uparrow}$ increases faster for SGR compared to RR. Given the relatively low precision and recall scores for RR and the higher scores for our method, one can conclude that RR underestimates the uncertainty boundaries, while this problem is largely reduced for the uncertainty boundaries estimated by SGR. 

\section{Conclusion and Discussion}

We have introduced a novel method for segmentation-guided MRI reconstruction and have shown that it can generate meaningfully diverse reconstructions that lead to upper and lower bound segmentations. These segmentation can offer an intuitive way of understanding the segmentation uncertainty that is caused by the ill-posed nature of the reconstruction problem. As our method also offers more reliable uncertainty boundaries compared to the standard method of repeated reconstruction, we believe that it can be helpful in clinical practice, whenever a trustworthy uncertainty estimation is required.

\begin{table}[t]
\centering
\caption{Quantitative results for experiments for different accelerations. All reported segmentation metrics are calculated on the 3D-volumes of all 4 segmentation classes. Median values are reported. SSIM and PSNR values are calculated as the average over all reconstructed volumes. Note that it is not clear if larger or smaller ratios of $V_{unc}/V^{\uparrow}$ are preferable, due to the inherent uncertainty of the reconstruction problem.}
\label{tab:results}
\begin{tabular}{llccccc}\toprule
& & \multicolumn{3}{c}{Segmentation} & \multicolumn{2}{c}{Reconstruction}
\\\cmidrule(lr){3-5}\cmidrule(lr){6-7}
Acc. & Method & Prec. $S^{\downarrow}$  & Recall $S^{\uparrow}$ & $V_{unc}/V^{\uparrow}$ & SSIM  & PSNR\\\midrule
\multirow{2}{*}{4x} & SGR (ours) & $\textbf{0.992}$ & $\textbf{0.995}$ & $0.126$  & $0.914$ & $\textbf{34.6}$ \\
 & RR &0.982 & 0.989 & 0.045 & \textbf{0.915} & \textbf{34.6} \\\midrule
\multirow{2}{*}{8x} &SGR (ours) & \textbf{0.989} & \textbf{0.992} & 0.322 & \textbf{0.857} & 30.8\\
  & RR & 0.971 & 0.976 & 0.075 & \textbf{0.857} & \textbf{30.9} \\\midrule
  \multirow{2}{*}{12x} & SGR (ours) & \textbf{0.984} & \textbf{0.991} & 0.473 & 0.818 & \textbf{29.1} \\
  & RR & 0.961 & 0.966 & 0.096 & \textbf{0.820} & \textbf{29.1} \\\midrule
  \multirow{2}{*}{16x} & SGR (ours) & \textbf{0.980} & \textbf{0.990} & 0.586 & 0.788 & 27.9 \\
  & RR & 0.951 & 0.957 & 0.113 & \textbf{0.789} & \textbf{28.0} \\\bottomrule
\end{tabular}
\end{table}

In future work, we want to analyze if robust segmentation networks introduced in \cite{naman_robust_seg} can help to reduce $V_{unc}$ while maintaining high precision and recall scores. In this work, however, we have focused on more commonly used neural networks, to show the easy applicability of our method for existing pipelines.

\begin{credits}
\subsubsection{\ackname} Funded by the Deutsche Forschungsgemeinschaft (DFG, German Research Foundation)
under Germany’s Excellence Strategy – EXC number 2064/1 – Project number 390727645.
The authors thank the International Max Planck Research School for Intelligent Systems
(IMPRS-IS) for supporting Jan Nikolas Morshuis.

\subsubsection{\discintname} The authors have no competing interests to declare.
\end{credits}

\newpage

\bibliographystyle{splncs04}
\bibliography{main}
\newpage
\appendix

\section{Supplementary Material}

\subsection{Ablation study for the weighting hyperparameter $b$}

\begin{figure}
    \centering
    \includegraphics[width=\textwidth]{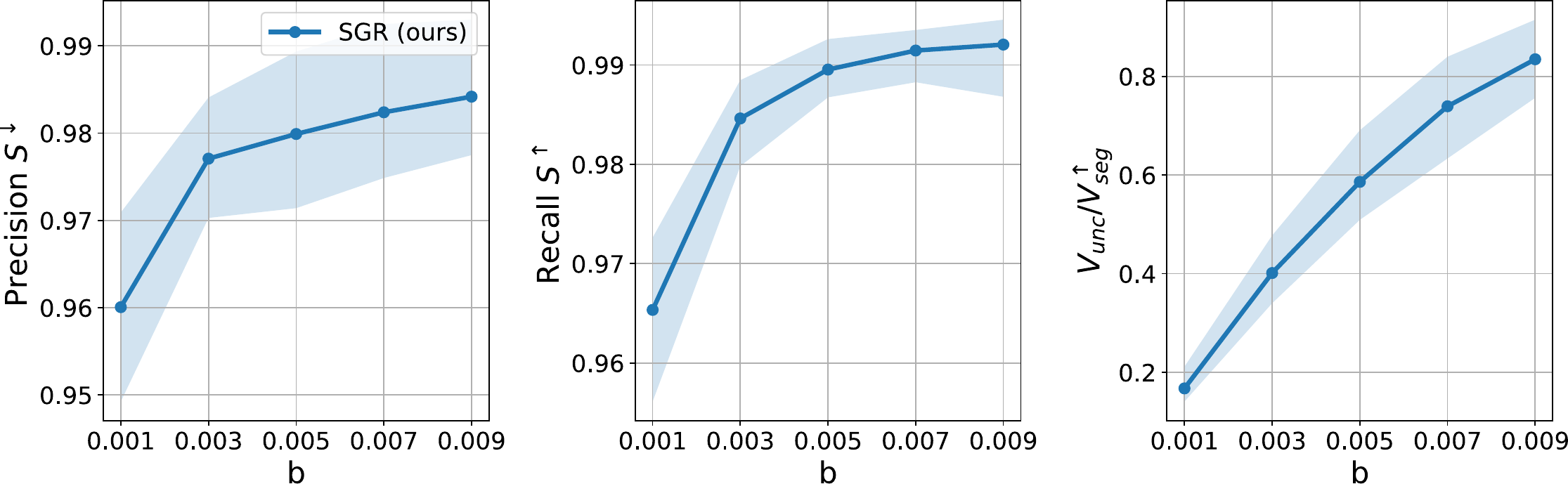}
    \caption{Ablation study with 16x acceleration on the hyperparameter $b$, which is responsible to weight the influence of $\nabla_{x_t} \mathcal{L}_m^{\uparrow / \downarrow}$ in the calculation of $x_{t-1}$ (see Eqs. 13 and 14 in the main paper). Note that this parameter largely influences the uncertainty volume $V_{unc}$, as well as the Precision of $S^{\downarrow}$ and the Recall of $S^{\uparrow}$.}
    \label{fig:ablation_study}
\end{figure}

\end{document}